\documentclass[preprint,showpacs,preprintnumbers,amsmath,amssymb]{revtex4}
\usepackage{amsmath}
\usepackage{graphicx}
\usepackage{amssymb}
\usepackage{bm}
\usepackage{color}

\begin{document}

\title{Effects of the entrance channel and fission barrier
in synthesis of superheavy element $Z$=120}
\author{A. K. Nasirov$^{1,2}$,
G. Mandaglio$^3$,   G. Giardina$^3$, A. Sobiczewski$^4$, A. I. Muminov$^2$}

\affiliation{$^1$Joint Institute for Nuclear Research, Joliot-Curie 6, 141980 Dubna, Russia,\\
$^2$Institute of Nuclear Physics, Ulugbek, 100214, Tashkent, Uzbekistan,\\
$^3$Dipartimento di Fisica dell' Universit\`a di Messina, Salita Sperone 31, 98166 Messina,
and Istituto Nazionale di Fisica Nucleare, Sezione di Catania,  Italy,\\
$^4$Soltan Institute for Nuclear Studies, Hoza 69, PL-00-681 Warsaw, Poland}

\begin{abstract}
The fusion and evaporation residue cross sections for
the $^{50}$Ti+$^{249}$Cf and $^{54}$Cr+$^{248}$Cm reactions
calculated by the combined dinuclear system and advanced statistical
models are compared. These reactions are considered to be used
to synthesize the heaviest superheavy element.
 The $^{50}$Ti+$^{249}$Cf reaction is
more mass asymmetric than $^{54}$Cr+$^{248}$Cm
and the fusion excitation function for the former reaction is higher than  the
one for the latter reaction. The evaporation
 residue excitation functions for the mass asymmetric reaction is
 higher in comparison with the one of the $^{54}$Cr+$^{248}$Cm reaction.
  The use of the mass values of superheavy nuclei calculated in the
 framework of the macroscopic-microscopic model by  the Warsaw
 group  leads to smaller evaporation residue cross section
 for both the reactions in comparison with the case of using the masses
 calculated by  Peter  M\"oller  {\it et al}. The $^{50}$Ti+$^{249}$Cf
 reaction is more favorable in comparison with the $^{54}$Cr+$^{248}$Cm
 reaction: the maximum values of the excitation function
 of the 3n-channel of the evaporation residue formation
 for the $^{50}$Ti+$^{249}$Cf and $^{54}$Cr+$^{248}$Cm reactions are about
 0.1 and 0.07 pb, respectively, but the yield of the 4n-channel
 for the former reaction is lower (0.004 pb) in comparison with the one
 (0.01 pb) for the latter reaction.
\end{abstract}

\date{Today}
\maketitle

\section{Introduction}
\label{sec:1}

The  synthesis of  the superheavy elements with  $Z=$
114--118  by hot-fusion reactions of $^{48}$Ca
 with actinide targets \cite{Oganessian04,FLNR}
and   with Z=110, 111, and 112 by using  cold-fusion reactions
\cite{HofMuen,RIKEN} with lead- and bismuth-based targets (shell
closed spherical nuclei) have been reported.
The cross section of the evaporation residue (ER) formation being a
 superheavy element is very small: some picobarns, or even the part of
picobarn.
The lightest isotope $^{278}$113 of the superheavy element $Z$=113
which was synthesized in the cold-fusion $^{70}$Zn+$^{209}$Bi reaction
was observed with a cross section value equal to some percents
of picobarn \cite{RIKEN}.

To find favorable reactions (projectile and target pair) and the
optimal beam  energy range leading to larger cross sections of
synthesis of superheavy elements, we should establish conditions
leading to increase  as much as possible the events
of ER formation. The ER formation process is often considered as
the third stage of the reaction mechanism in heavy ion collisions
at near the Coulomb barrier energies.
  The first stage is a capture--formation of the dinuclear system (DNS) after
full momentum transfer of the relative motion of colliding nuclei
into the shape deformation, excitation energy and rotational
energy of nuclei. The capture takes place if the initial energy of
projectile in the center-of-mass system is
sufficiently large to overcome the interaction barrier (Coulomb
barrier + rotational energy of the entrance channel)
 and
it is dissipated leading to trap DNS into the well of the
nucleus-nucleus interaction potential \cite{GiaEur2000}. The same
mechanism takes place in both kinds of reactions, but the
probability of the realization of each stage of the whole
mechanism  is different in cold and hot fusion
reactions \cite{ColdHotMess}.

We calculate the cross section of ER formed after each step $x$
of the de-excitation cascade after the emission from the hot CN
of particles $\nu(x)$n + $y(x)$p + $k(x)\alpha$ + $s(x)$
(where $\nu(x)$, $y$, $k$, and $s$ are numbers of neutrons, protons,
$\alpha$-particles, and $\gamma$-quanta) by formula
(See Refs. \cite{GiaEur2000,FazioEPJ2004}):
\begin{equation}
\sigma_{\rm ER}(E_{\rm c.m.})=\Sigma^{\ell_d}_{\ell=0}
(2\ell+1)\sigma^{(x-1)}_{\ell}(E_{\rm c.m.})W^{(x-1)}_{\rm sur}(E_{\rm c.m.}+Q_{\rm gg},\ell),
\end{equation}
where $\sigma^{(x-1)}_{\ell}$ is the partial formation cross section
of the excited intermediate nucleus of the $(x-1)$th step and
$W^{(x-1)}_{\rm sur}$ is the survival probability of the
$(x-1)$th intermediate nucleus against fission along the
de-excitation cascade of CN.
It is clear that the first de-excitation step occurs
with the compound nucleus which is formed at complete fusion:
\begin{equation}
\sigma^{(0)}_{\ell}(E_{\rm c.m.},\ell)
 =\sigma_{\rm fus}(E_{\rm c.m.},\ell).
\end{equation}
The fusion cross section is related to the number of events
corresponding to the transformation of the dinuclear system into
compound nucleus in competition with the quasifission process.
It is defined by the product of the partial capture cross section
and the related fusion factor PCN which allows to take into account
the competition between the complete fusion and quasifission processes
(See Refs.  \cite{VolPLB1995,AdamianPRC2003}):
\begin{equation}
\label{Eq3}
\sigma_{\rm fus}(E_{\rm c.m.},\ell)=
\sigma_{\rm capture}(E_{\rm c.m.},\ell)P_{\rm CN}(E_{\rm c.m.},\ell).
\end{equation}
Our method of calculation
(also including the advanced statistical method \cite{ArrigoPRC1992,ArrigoPRC1994,SagJPG1998})
 of the ER cross sections takes into account the damping of
the shell correction in the fission barrier as a function
of the excitation energy and orbital angular momentum.
This is accounted for the various steps of the de-excitation
cascade of the compound nucleus leading to the fission fragments
or the ER nuclei in the exit channel \cite{FazioEPJ2004,FazioJPSJ72,FazioJPSJ74}.

The study of dynamics of these
processes in heavy ion collisions at near the Coulomb barrier
energies showed that complete fusion does not occurs immediately
in the case of  massive nuclei collisions
\cite{Back32,Dvorak,VolPLB1995,AdamianPRC2003,FazioEPJ2004}. The
quasifission process competes with formation of compound nucleus
(CN). This  process occurs when DNS prefers to
break down into two fragments instead of to be transformed into
fully equilibriated  CN. The number of events going to
quasifission increases drastically by increasing  the sum of the
Coulomb interaction and rotational energy in the entrance channel
\cite{GiaEur2000,FazioPRC2005}. The Coulomb interaction increases
 by increasing the charge number of the projectile or target nucleus,
as well as it increases  by decreasing   the charge asymmetry of
colliding nuclei at fixed total charge number of DNS.

\begin{figure}
\vspace*{2.0cm}
\begin{center}
\resizebox{0.80\textwidth}{!}{\includegraphics{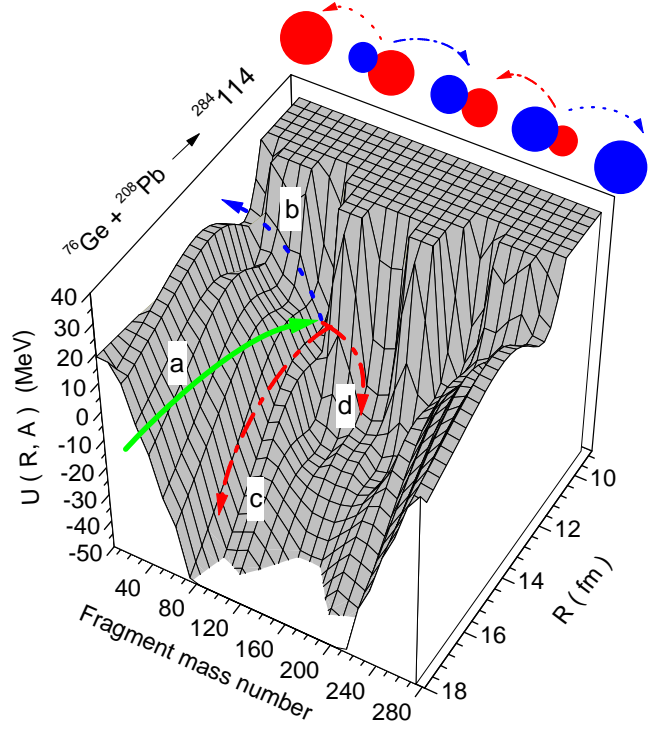}}
\vspace*{-3.60 cm} \caption{\label{DNSGePb} (Color online) Potential energy surface
 calculated for the DNS leading to formation of the  $^{284}$114 compound  nucleus
 as a function of the relative distance between the centers of mass
of interacting nuclei and mass number of a fragment.}
\end{center}
\end{figure}

Another reason decreasing the yield of ER is the fission of a
heated and rotating CN which is formed in
competition with quasifission. The stability of massive CN
decreases due to the decrease of the fission barrier by increasing
its excitation energy $E^*_{\rm CN}$ and angular momentum $L$
\cite{ArrigoPRC1992,ArrigoPRC1994,SagJPG1998}. The stability of
the transfermium nuclei is connected with the red
appearance of the shell correction in their binding energy
\cite{Sobiczewski2007}, which  is sensitive to the
 angular momentum and $E^*_{\rm CN}$ values. The
fusion-fission takes place when the compound nucleus cannot
survive against fission due to smallness of its fission barrier
which decreases  by increasing the excitation
energy $E^*_{\rm CN}$ and/or  angular momentum $\ell_{\rm CN}$.
In the cold fusion reactions
the desired flow of nucleons from the projectile-nucleus to the
target-nucleus (in this case $^{208}$Pb or $^{209}$Bi) is strongly hindered
when the projectile is heavier than $^{70}$Zn. This is connected with
the dependence of the potential energy surface (PES) on the mass and charge
asymmetries and on the shell effects in the binding energies of
colliding nuclei (see Fig. \ref{DNSGePb}).
The use of nuclear binding energies including shell effects in calculations
of the PES and driving potential of DNS leads to the appearance of hollows
on the PES around the charge and mass symmetries corresponding to the constituents of
DNS with the magic proton or/and neutron numbers
 (see Figs. 4 and 5 in Ref. \cite{GiaEur2000}).

 The charge asymmetry of the entrance channel for the ``cold fusion''
reactions is placed on the hollow between the Businaro-Gallone point
(b) in Fig. \ref{DNSGePb} and
the valley of the charge symmetric channel (point (d) in Fig. \ref{DNSGePb}).
 The intrinsic fusion barrier $B^*_{\rm fus}$ increases by increasing the
projectile charge and mass numbers. It is determined as the difference
of values of the potential energy surface on the point where DNS
had been captured (on the bottom of the potential well
of the nucleus-nucleus interaction considered as a function
of the relative distance $R$ between the centers of nuclei)
and on the ``saddle point'' in the fusion valley (near point ``b'' of Fig. \ref{DNSGePb})
(for details see Refs. \cite{GiaEur2000,NasirovNPA759,NasirovPRC79}).
This  fact leads to a strong increase of the hindrance to complete
fusion and the probability of compound nucleus formation becomes
 very small.

 The superheavy elements $Z$=110, 111, 112, and 113 were synthesized
 in  ``cold fusion'' reactions by bombarding
$^{208}$Pb and $^{209}$Bi nuclei which have $N=126$ neutrons. The
cold fusion reactions were preferable at synthesis of superheavy
elements up to $Z$=112. For example, the maximum value of ER cross
section ($\sigma_{\rm ER}$) at synthesis of the superheavy element
$^{265}_{108}$Hs  in the cold fusion reaction $^{58}$Fe+$^{208}$Pb
\cite{SHof108} was $\sigma_{\rm ER}$=65 pb. This value is
 about one order of magnitude higher than  the ER
cross section $\sigma_{\rm ER}$=7 pb measured in the hot fusion
reaction $^{28}$Si+$^{238}$U \cite{Dvorak} , but  the synthesis of
superheavy elements becomes more favorable in hot fusion reactions
starting from $Z$=112.

Therefore, all of the last group of elements  with  $Z=$114, 115, 116, 117 and 118 were
synthesized in the hot fusion reactions where the actinide targets
 $^{242,244}$Pu, $^{243}$Am, $^{248}$Cm, $^{249}$Bk, and $^{249}$Cf were
bombarded by the neutron rich isotope
$^{48}$Ca.

\section{Advantage of hot fusion reactions with massive nuclei}

The  advantage of  hot fusion reactions in comparison with cold fusion
reactions is connected with the relatively small hindrance in the
compound nucleus formation.  Because  the charge asymmetry of the
entrance channel ($^{48}$Ca) in hot fusion reactions
is placed closer to the
Businaro-Gallone point (see Fig. \ref{DNSGePb}),
consequently, the intrinsic fusion barrier $B^*_{\rm fus}$  of DNS
is smaller in comparison with  the one for cold fusion reactions
 ($^{76}$Ge).
The  large excitation energy of compound
nucleus is an inevitable circumstance in the hot fusion
 reactions because after capture and formation of the DNS, the value of PES
 corresponding to the entrance channel charge asymmetry is settled at  higher
 points of its hollow in comparison with the case of cold fusion reactions.
 Therefore, even  if the compound nucleus is formed by the minimum possible
energy beam, it is excited at energies  higher than 30 MeV. As an example,
 to show such a  strong difference of the hindrance to complete fusion,
we compare in Table I the values of fusion probability ($P_{\rm CN}$) for two
sets of the cold  and hot  fusion reactions.

\begin{table}[hpt]
\label{ColdHot}
\caption{Comparison of the fusion probabilities ($P_{\rm CN}$)
for the cold (left side) and hot (right side) fusion reactions
calculated in the dinuclear system model \cite{GiaEur2000,NasirovNPA759,NasirovPRC79}}. \\
\begin{tabular}{|c|c|c|c|c|c|c|c|c}
  \hline

      Cold fusion  & $Z_{\rm CN}$ &$\eta=\frac{A_2-A_1}{A_1+A_2}$ & \hspace*{0.5cm}$P_{\rm CN}\cdot 10^{-8}$ \hspace*{0.5cm} & Hot fusion & $Z_{\rm CN}$  & $\eta=\frac{A_2-A_1}{A_1+A_2}$&
     $\hspace*{0.5cm} P_{\rm CN}\cdot 10^{-2}$ \hspace*{0.5cm}\\
           reactions     &              &    &                      &   reactions         &      &&   \\
  \hline
     $^{64}$Ni+$^{208}$Pb$^*$ & 110 & 0.529 & 14.0  & $^{48}$Ca+$^{243}$Am$^{\dag}$ & 115 & 0.670 & 5.02  \\
   $^{64}$Ni+$^{209}$Bi$^*$ & 111 & 0.531 & 7.0  & $^{48}$Ca+$^{248}$Cm$^{\dag}$ & 116 & 0.676 & 1.13   \\
   $^{70}$Zn+$^{208}$Pb$^*$ & 112 & 0.496 & 0.25  & $^{48}$Ca+$^{249}$Bk$^{\ddag}$ & 117 & 0.677 & 2.06  \\
   $^{70}$Zn+$^{209}$Bi$^*$ & 113 & 0.498 & 0.052  & $^{50}$Ti+$^{249}$Cf$^{\ddag}$ & 120 & 0.666 & 0.112   \\
   $^{76}$Ge+$^{208}$Pb$^*$ & 114 & 0.465 & 0.012  & $^{54}$Cr+$^{248}$Cm$^{\ddag}$ & 120 & 0.642 & 0.0231   \\
  \hline
    \end{tabular}

\hspace*{-5.6cm} $^*$   The estimations made from the results of Ref. \cite{GiaEur2000}.
\\
\hspace*{-5.6cm}
  $^{\dag}$  The estimations made from the results of Ref. \cite{FazioEPJ2004}.
\\
\hspace*{-9.2cm}  $^{\ddag}$ The estimation of this work.\\
\end{table}

The small cross section of the ER formation in hot fusion
reactions is connected with the small survival probability against
fission ($W_{\rm surv}\approx 10^{-8}$) of the heated and rotating compound
nucleus. The synthesis of superheavy elements with $Z$=117 and 118 at
the Flerov Laboratory of Nuclear Reactions of JINR in Dubna
(Russia), as well as the confirmation of the Dubna group's results
for the new elements with $Z$=114 and 116  at the Lawrence
Berkeley National Laboratory (USA)  \cite{Stavsetra} and  at GSI
(Darmstadt, Germany) \cite{SHofmann116} by the SHIP group, caused
the new attempts to reach  a heavier element, $Z$=120. Theoretical
estimations of the ER cross sections
 for the $^{54}$Cr+$^{248}$Cm, $^{58}$Fe+$^{244}$Pu,
and $^{64}$Ni+$^{238}$U reactions
have already been made  in Refs.\cite{LiuBao120,Zagreb117,AdamAnt,NasirovPRC79}.
 In the experiment with the $^{58}$Fe+$^{244}$Pu reaction,
 which was reported in Ref. \cite{Ogan120},
   no event  for the synthesis of the $Z=120$ element was  observed:
  the upper   limit  of cross section of 0.4 pb at
 $E^*_{\rm CN}$=46.7 MeV  was estimated.
The results of two experiments  at GSI (Darmstadt),
 where the $^{64}$Ni+$^{238}$U reaction was used, did not show events  for the synthesis
 of  the $Z=120$ element.
 The  $^{54}$Cr+$^{248}$Cm reaction, which seems to be  the more favorable
 among the above-mentioned reactions,  was recently  performed  at GSI.

\begin{table}[pt]
\label{Comp120ThMod}
\caption{Comparison of the predicted maximum values of the evaporation residues
cross section ($\sigma_{\rm ER}$)
in the $^{54}$Cr+$^{248}$Cm and $^{50}$Ti+$^{249}$Cf reactions
obtained in Refs.\cite{Zagreb117,KSiwek,LiuBao120}
with our results  for the 3  and 4 neutrons emission channels as a function of the
collision energy in the center-of-mass system $E_{\rm c.m.}$.
The presented data about maximum values from Refs.\cite{Zagreb117,KSiwek,LiuBao120} were extracted
from the figures of the ER excitation functions.
}
\vspace{0.5cm}
{\begin{tabular}{|cccc|cccc|c|}
  \hline
    &$^{50}$Ti+$^{249}$Cf &&&&&  $^{54}$Cr+$^{248}$Cm && Reference   \\
  \hline
$E_{\rm c.m.}$ & $\sigma^{(\rm 3n)}_{\rm ER}$ &$E_{\rm c.m.}$ & $\sigma^{(\rm 4n)}_{\rm ER}$ & $E_{\rm c.m.}$
& $\sigma^{(\rm 3n)}_{\rm ER}$ & ~$E_{\rm c.m.}$ &  $\sigma^{(\rm 4n)}_{\rm ER}$ &  \\
MeV & fb & MeV & fb & MeV & fb & MeV & fb &\\
  \hline
236.0 & 1.5 & - & - & 248.2  & 0.2  & - & - & \cite{AdamAnt}$^*$\\
236.0 & 40.0 & 241.0 & 46.0 & 246.7  & 14.0  & 249.6 & 28.0 & \cite{Zagreb117}$^*$\\
231.5 & 60.0 & 232.5 & 40.0  & - & - & - & - &  \cite{LiuBao120}$^*$\\
227.5 & 760.0 & 239.0 & 28.0 & 241.5 & 76.0  & 252.0& 12.0 &  \cite{KSiwek}$^{\dag}$\\
225.0 & 10.0 & 231.5 & 2.5 & 237.2 & 55.0 & 241.0 & 13.0 & This work$^*$ \\
  \hline
\end{tabular}}
\\
\hspace*{-1.1cm}$^*$ The corresponding authors used data from the mass
table presented in Ref. \cite{MolNix}.
\\
\hspace*{-0.9cm}$^{\dag}$ The corresponding authors used data from the mass table presented in
 Ref. \cite{Muntian03}.
 \\

\end{table}

 In Table \ref{Comp120ThMod}, the predictions of the maximum values of the
 evaporation residues excitation functions for the 3n- and 4n-channel
 by  different models (see Refs.\cite{LiuBao120,Zagreb117,AdamAnt,KSiwek}
 are presented. The results presented in Refs. \cite{LiuBao120,Zagreb117,AdamAnt}
 were obtained by using the theoretical binding energies from the mass table
 by P. M\"oller {\it et al.} \cite{MolNix}, while the authors of Ref. \cite{KSiwek}
  have used the mass data calculated by I.
  Muntian {\it et al.} \cite{Muntian03}.

The difference between compared results in Table \ref{Comp120ThMod}  can be explained
by  three  main reasons:
1) the authors used different methods to estimate the formation probability  of
the heated and rotating compound nuclei $^{299}$120 and $^{302}$120 in the $^{50}$Ti+$^{249}$Cf
and $^{54}$Cr+$^{248}$Cm reactions (details of calculations can be found in the corresponding
references); 2) the survival probability calculations of the compound
 nucleus against fission  are sensitive to  the values of the statistical model
 parameters; 3) the use of different theoretical  nuclear mass tables
 can give relevant difference in  the  values of  nuclear binding energy.

\begin{figure}
\vspace*{-1.0cm}
\begin{center}
\resizebox{0.80\textwidth}{!}{\includegraphics{{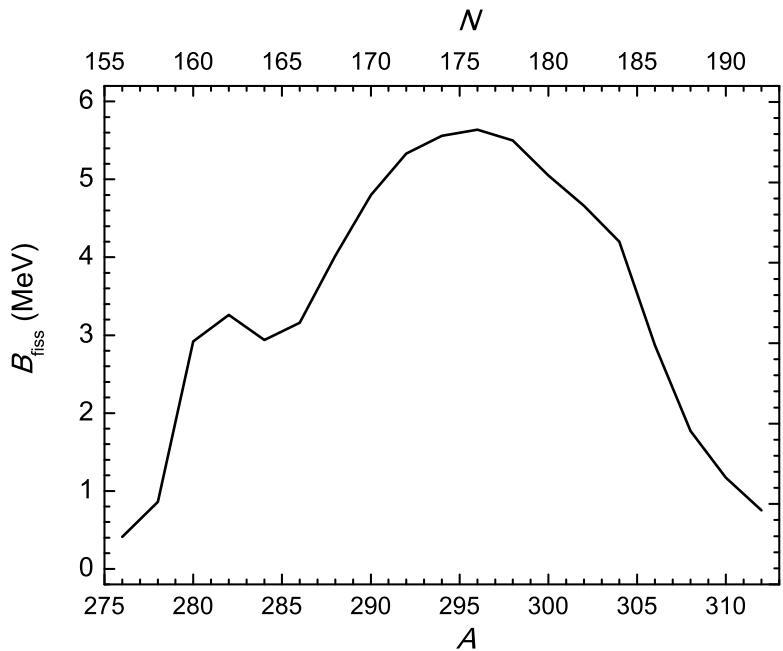}}}
\vspace*{-0.60 cm} \caption{\label{Fissbarr} The fission barriers
for the isotopes of superheavy element $Z$=120 calculated by the
macroscopic-microscopic model  of Ref.
\cite{KowalPRC}.}
\end{center}
\end{figure}

 The theoretical results obtained by  the Warsaw group
 within the macroscopic-microscopic model \cite{KowalIJMP,KowalPRC} showed
 the increase of the fission barrier of the isotopes of the superheavy element
 $Z$=120 at decreasing its mass number  from the value $A$=310 down to $A$=296
(see Fig. \ref{Fissbarr}).
 This effect was obtained   taking into account non-axial quadrupole deformation.
 Therefore,  we estimated in this paper the ER cross sections
 for the $^{50}$Ti+$^{249}$Cf reaction
 leading to formation of the isotope $A$=299 of the  $Z$=120  element to observe the
  effect of the  increasing  barrier  on the ER formation.
 In Section \ref{CompCapFus}, we compare  our results of capture, fusion
 and evaporation residue cross sections for
 the $^{50}$Ti+$^{249}$Cf and $^{54}$Cr+$^{248}$Cm reactions
 to find out the role of the entrance channel and  fission barriers on the reaction products.

\section{capture, fusion, and evaporation residue cross sections for
  the $^{50}{\rm Ti}$+$^{249}{\rm Cf}$ and $^{54}{\rm Cr}$+$^{248}{\rm Cm}$ reactions}\label{CompCapFus}

\begin{figure}
\vspace*{-1.0cm}
\begin{center}
\resizebox{0.80\textwidth}{!}{\includegraphics{{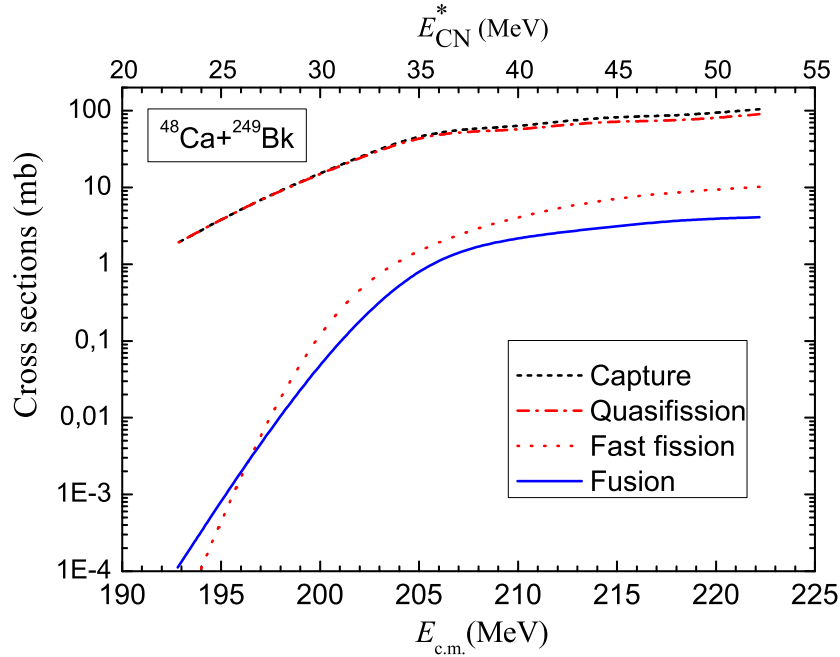}}}
\vspace*{-1.00 cm} \caption{\label{CapFusCaBk} (Color online) Capture (dashed line),
quasifission (dot-dashed line),  fast fission (dotted line) and fusion (solid line)
cross sections calculated by the DNS model for the $^{48}$Ca+$^{249}$Bk reaction.
 The excitation energy $E^*_{\rm CN}$ of compound nucleus (top axis)
is calculated by the use of the  M\"oller and Nix mass table \cite{MolNix}.}
\end{center}
\end{figure}

 The calculations of capture and fusion cross sections were performed in the framework
 of the DNS model.  The details of this model can be found in Refs.
 \cite{FazioEPJ2004,FazioMPL2005,NasirovNPA759,NasirovPRC79}. The partial fusion
 cross sections $\sigma_{\rm fus}^{(\ell)}$ obtained in the DNS model were used
 to calculate evaporation residue cross sections by the advanced statistical model
 \cite{ArrigoPRC1994,SagJPG1998}.
 We have described the experimental data \cite{Ogan117} of the ER cross section
 for the $^{48}$Ca+$^{249}$Bk  reaction leading to the superheavy element  $Z$=117.
 The results of calculations for the capture and fusion cross sections for the
 $^{48}$Ca+$^{249}$Bk  reactions are presented in Fig. \ref{CapFusCaBk}.

The capture of projectile by target at a given beam energy and  for all possible orbital angular
momentum  values is determined as trapping of the system into potential well
 of the nucleus-nucleus
interaction after full momentum transfer and dissipation of the relative kinetic energy
into the deformation  and excitation energy of nuclei (for details see Refs.
\cite{FazioMPL2005,NasirovNPA759,FazioPRC2005}. The number of the partial waves contributing
to the capture cross section is found by solution of the classical equation of motion for the
relative distance between centers of the interacting nuclei and angular momentum \cite{NasirovNPA759}.
The friction coefficients are calculated  by using the expression obtained by
averaging coupling term between intrinsic excitation in nuclei and nucleon exchange
between them \cite{AdamianFric}.
 \begin{figure}
\vspace*{2.0cm}
\begin{center}
\resizebox{1.0\textwidth}{!}{\includegraphics{{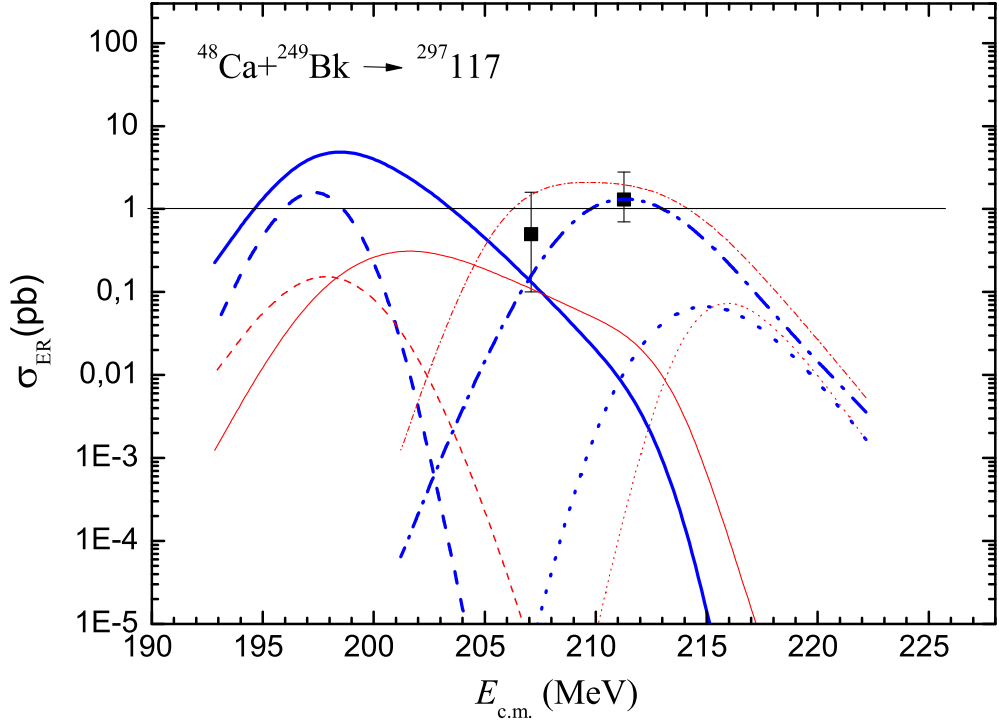}}}
\vspace*{-0.90 cm} \caption{\label{ER117} (Color online)
Comparison between the evaporation residue excitation functions
for the $^{48}$Ca+$^{249}$Bk reaction calculated by using mass
tables of  M\"oller and Nix \cite{MolNix} (thick
lines) and  of the Warsaw group \cite{Muntian03}
(thin lines) for the 2n (dashed lines), 3n (solid lines), 4n
(dot-dashed lines), and 5n (dotted lines) channels calculated by
the advanced statistical model
\cite{ArrigoPRC1992,ArrigoPRC1994,SagJPG1998}. The experimental
data  of Ref. \cite{Ogan117} are presented
 by squares.}
\end{center}
\end{figure}

 One can see in Fig. \ref{CapFusCaBk} that the hindrance to fusion increases at
lower energies  $E_{\rm c.m.}<$205 MeV because  at these low energies the collisions
 with small orientation angles ($\alpha_P$-projectile and $\alpha_T$-target) of the axial
 symmetry axes of colliding nuclei relative to the beam direction \cite{NasirovNPA759,HindePCR74}
  can  only contribute. At capture  of colliding nuclei
 with  small   orientation angles $\alpha_P$ and $\alpha_T$,
the intrinsic barrier $B^*_{\rm fus}$ for the
  transformation into  the compound nucleus  is large \cite{NasirovNPA759}.
 Therefore, at energies  $E_{\rm c.m.}<$205 MeV the capture of projectile by target-nucleus in
collisions with large orientation angles $\alpha_P$ and $\alpha_T$
 is impossible: the initial collision energy is not
 sufficient to overcome the Coulomb barrier which is large in comparison with  the one
 in the case of  small orientation angles.
So, the hindrance to complete fusion depends on the orientation angles:
 more elongated shape of the DNS formed  at collisions with small
 orientation angles (tip-to-tip  configurations)  promotes  the  quasifission  rather than
 the formation of the compound nucleus \cite{NasirovNPA759,HindePCR74}.
  Therefore,  a sufficiently high collision energy
 $E_{\rm c.m.}$ (as compared with the Bass barrier)
 was chosen in the experiments aiming in the synthesis of superheavy
 elements in ``hot fusion'' reactions with $^{48}$Ca on the actinide nuclei Pu, Am,
 Cm, Bk, and Cf  with the purpose of including the contributions of large orientation
 angles of the axial symmetry of the target nucleus.

 Theoretical results of the ER cross  sections for the synthesis of the element $Z=117$ are
 compared  with experiment in Fig. \ref{ER117}.
In this figure, the  full squares show experimental data of the ER
cross  sections measured for the $^{48}$Ca+$^{249}$Bk reaction in
Ref. \cite{Ogan117};
 the curves show theoretical results obtained in this work for the 2n-(dashed line),
3n-(solid line), 4n-(dot-dashed line) and 5n-channel (dotted line)
 by the DNS and advanced statistical models by using  the mass tables
 of M\"oller and Nix \cite{MolNix} (thick lines) and  of Muntian {\it et al.} \cite{Muntian03} (thin lines).
 According to our results, $\sigma_{\rm ER}$ is larger at the collision energies around
 $E_{\rm c.m.}=$200--205  MeV. The survival probability $W_{\rm surv}$ of the heated
 compound nucleus increases  with the decrease of its excitation energy.

  The main  scope of this work is to reproduce the measured data
 for the superheavy element $Z=117$ and to make predictions for  $\sigma_{\rm ER}$ in the
 $^{54}$Cr+$^{248}$Cm and $^{50}$Ti+$^{249}$Cf reactions which can
 be used in the nearly future experiments.

\begin{figure}
\vspace*{3.0cm}
\begin{center}
\resizebox{0.80\textwidth}{!}{\includegraphics{{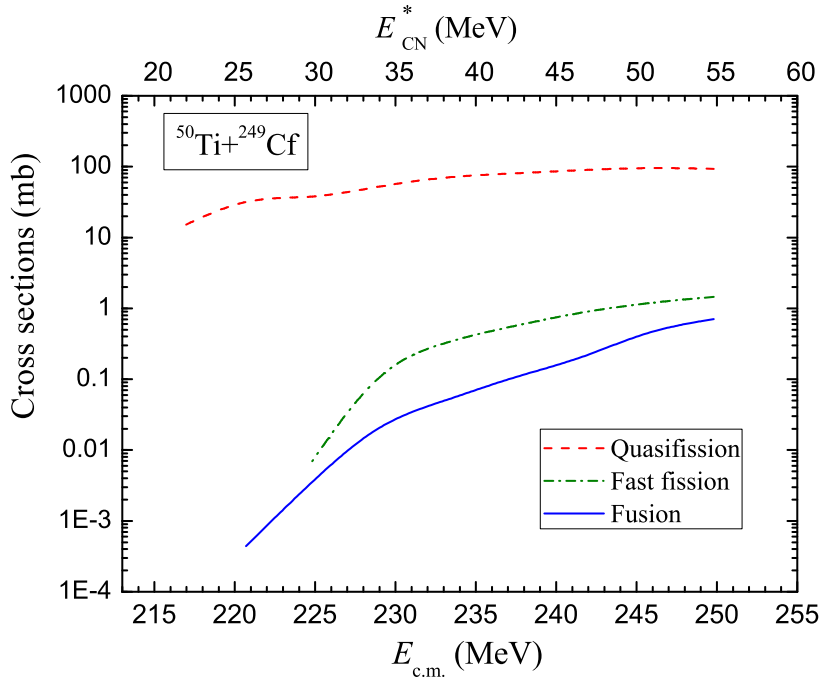}}}
\vspace*{-0.60 cm} \caption{\label{CapFusTiCf} (Color online) Quasifission (dashed line),
fast fission (dot-dashed line), and complete fusion (solid line) excitation functions
calculated by the DNS model \cite{FazioMPL2005,NasirovNPA759,FazioPRC2005} for the
$^{50}$Ti+$^{252}$Cf reaction  which could lead to the $^{299}120$ compound nucleus.
 The capture cross section is not shown here because it is completely overlapped
  with the quasifission cross section.   The excitation energy $E^*_{\rm CN}$ of compound nucleus (top axis)
is calculated by the use of the  M\"oller and Nix mass table \cite{MolNix}.}
\end{center}
\end{figure}

\begin{figure}
\vspace*{3.0cm}
\begin{center}
\resizebox{0.80\textwidth}{!}{\includegraphics{{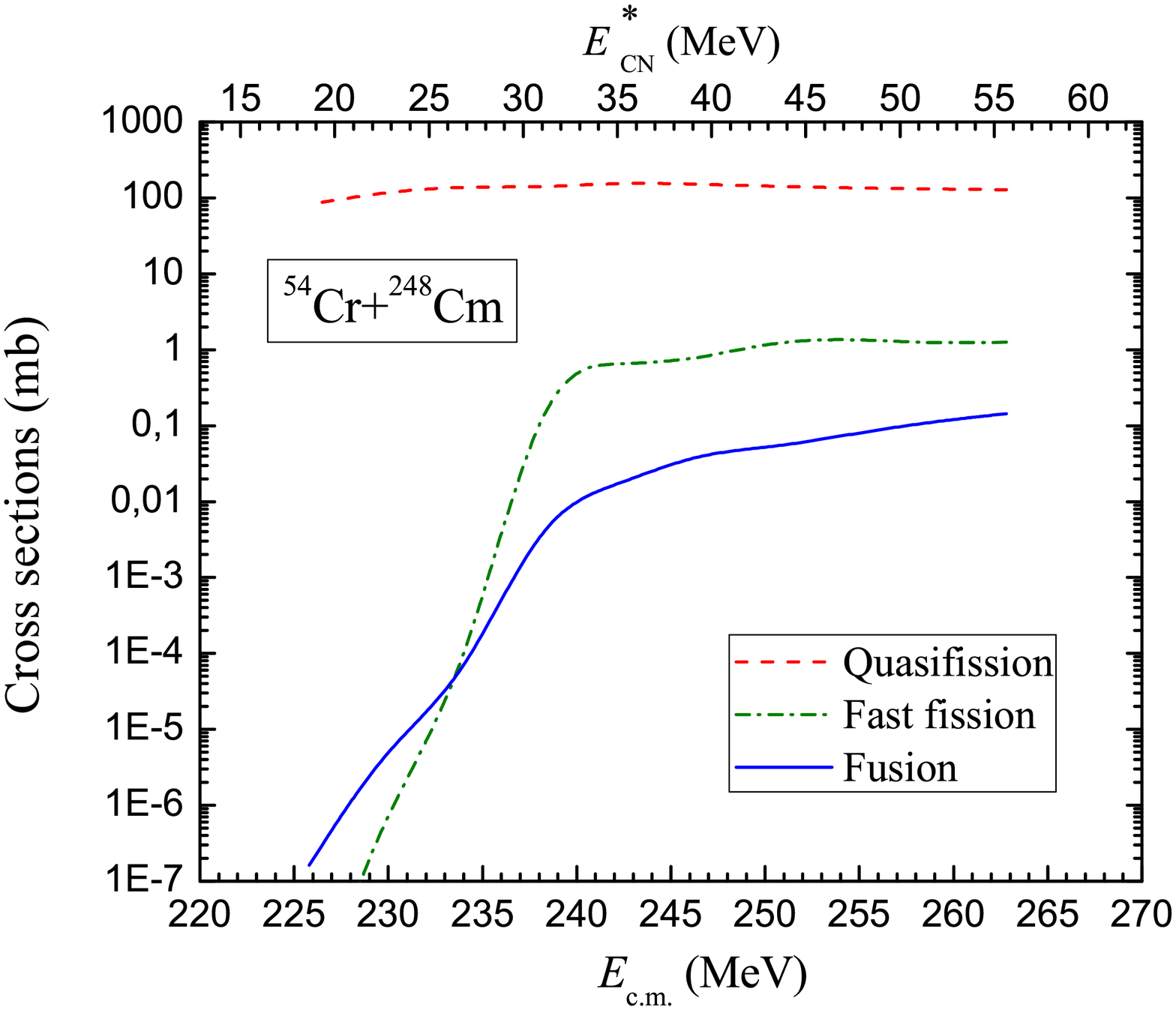}}}
\vspace*{-0.80 cm} \caption{\label{CapFusCrCm} (Color online)
The same as in Fig. \ref{CapFusTiCf} but for the $^{54}$Cr+$^{248}$Cm reaction
 which could lead to the $^{302}120$ compound nucleus.}
\end{center}
\end{figure}

 In Figs. \ref{CapFusTiCf} and \ref{CapFusCrCm} we present our theoretical results for
 quasifission, fast fission and complete fusion cross sections of the $^{50}$Ti+$^{249}$Cf
  and $^{54}$Cr+$^{248}$Cm  reactions. The capture cross section is not shown
  in Fig. \ref{CapFusTiCf} because it is completely overlapped
  with the quasifission cross section, since the sum of the fast fission and
  complete fusion is about 2--4 order of magnitude smaller than quasifission
  cross section.
  The comparison  between these figures show that, at low energies,
 the capture cross section
 in the $^{54}$Cr+$^{248}$Cm reaction is larger
than  that in the $^{50}$Ti+$^{249}$Cf
 reaction, while these cross sections become comparable  at
 larger energies.
  One can  also see in these figures that the fusion cross section is
  sufficiently larger for the $^{50}$Ti+$^{249}$Cf reaction in comparison with the one of
  the $^{54}$Cr+$^{248}$Cm reaction.
 The advance of the charge asymmetric system  appears at the second stage (fusion) of the
 reaction mechanism leading to formation of the evaporation residues.
 It is well known that the hindrance to complete fusion decreases by increasing the DNS
 charge asymmetry. At the same time the DNS quasifission barrier, $B_{\rm qf}$,
 increases because the Coulomb repulsion forces decrease  with the decrease of the product
 $Z_1\cdot Z_2$.
  Therefore, in spite of the fact that  the
  $^{50}$Ti+$^{249}$Cf system  has less neutrons in comparison with $^{54}$Cr+$^{248}$Cm,
  the probability of the compound nucleus formation is higher for the former reaction than
  for the latter  one. The more strong hindrance to complete fusion in  the case of the
  $^{54}$Cr+$^{248}$Cm reaction is connected with the larger intrinsic fusion barrier
  $B^*_{\rm fus}$ and smaller quasifission barrier $B_{\rm qf}$ for this reaction in
  comparison with $^{50}$Ti+$^{249}$Cf.

\begin{figure}
\vspace*{0.0cm}
\begin{center}
\resizebox{0.80\textwidth}{!}{\includegraphics{{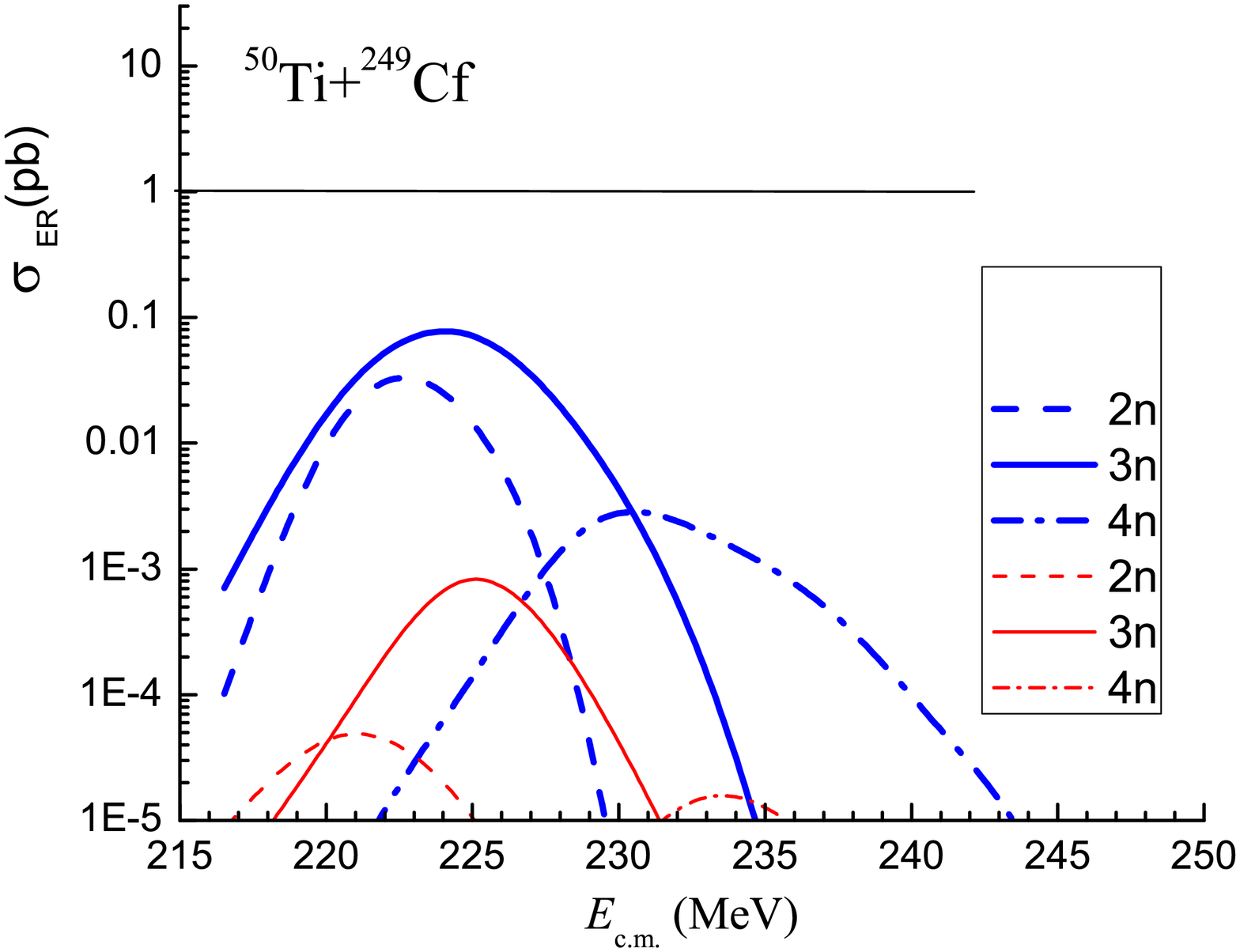}}}
\vspace*{-0.90 cm} \caption{\label{ERTiCf} (Color online)
Comparison between the evaporation residue
excitation functions for the $^{50}$Ti+$^{252}$Cf reaction
calculated by using mass tables of  M\"oller and Nix
\cite{MolNix} (thick lines) and of  the Warsaw group
\cite{Muntian03} (thin lines) for the 2n (dashed lines), 3n (solid
lines), 4n (dot-dashed lines), and 5n (dotted lines) channels
calculated by the advanced statistical model
\cite{ArrigoPRC1992,ArrigoPRC1994,SagJPG1998}. }
\end{center}
\end{figure}

\begin{figure}
\vspace*{0.0cm}
\begin{center}
\resizebox{0.80\textwidth}{!}{\includegraphics{{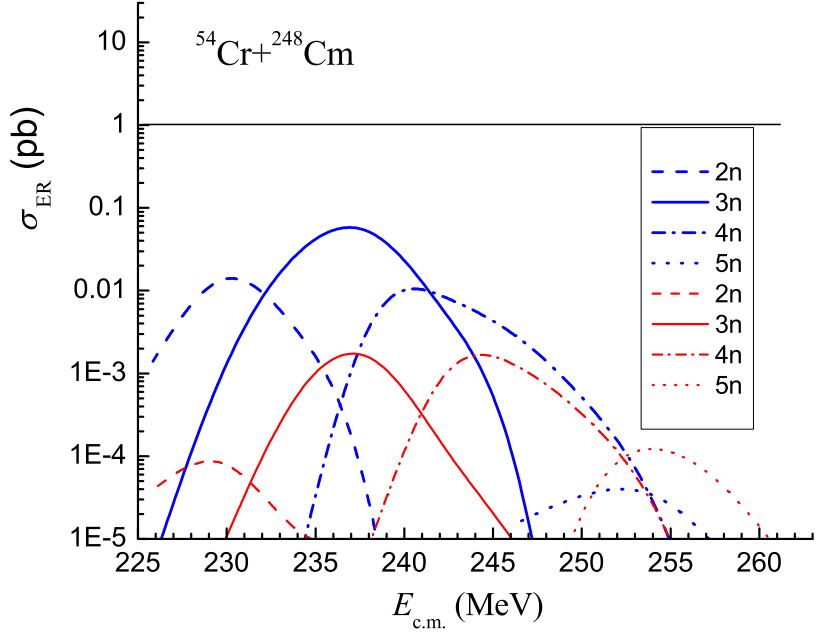}}}
\vspace*{-0.80 cm} \caption{\label{ERCrCm} (Color online)
The same as in Fig. \ref{ERTiCf} but for the $^{54}$Cr+$^{248}$Cm reaction.}
\end{center}
\end{figure}

 The theoretical excitation functions of evaporation residues which can be formed
 in different neutron-emission channels for these two systems are presented in Figs.
 \ref{ERTiCf} and \ref{ERCrCm}. In each of the figures the evaporation  residue
 cross  sections
  for the neutron-emission channels obtained by using  binding energies and fission barriers
  calculated in the microscopic-macroscopic  models of  M\"oller and Nix \cite{MolNix} and
 of the Warsaw group \cite{Muntian03} are compared. The difference between binding energies
 obtained by these two groups is in the range  of  2-3 MeV for the isotopes of superheavy
 nuclei  with $Z > 114$. This difference causes a difference between values
 of the branching ratios $\Gamma_n/\Gamma_f$  which are used in calculations
 of the survival probability of the heated and rotating nuclei.  The use of
 the binding energies \cite{Muntian03} and fission barriers \cite{KowalIJMP} of
 the Warsaw group  leads to two main consequences:
 the excitation energy of the compound nucleus will be lower because  the absolute
 value of $Q_{\rm gg}=B_{\rm proj}+B_{\rm targ}-B_{\rm CN}$ (negative) is larger:
$E^*_{\rm CN}=E_{\rm c.m.}+Q_{\rm gg}$,
 and  the fission probability of CN becomes higher in comparison with case of
 using fission barrier of the  M\"oller and Nix \cite{MolNix} model
 since taking into account triaxial deformations significantly reduces
 the fission barrier heights by up to 2.5 MeV for the $Z > 112 $ \cite{KowalIJMP}.

 Therefore, the evaporation residues
 cross sections obtained by the use of mass table calculated by the
 Nix-M\"oller microscopic-macroscopic  model  are  one order of magnitude larger
 in comparison with  the results obtained  by the use of  the mass table  of
  the Warsaw group.

 We should comment  on the difference between our  present results for the excitation function
 of evaporation residues $\sigma_{\rm ER}$ for the $x$n-channels in the $^{54}$Cr+$^{248}$Cm
 reaction  and  the ones given  in Ref. \cite{NasirovPRC79}:
 the  values of $\sigma_{\rm ER}$ presented in Fig. \ref{ERCrCm} are  much lower
 than those published in Ref. \cite{NasirovPRC79}. The analysis showed that
 the evolution of mass and charge distributions in the DNS constituents was
  very sensitive to the used  nuclear radius
 parameter $r_0$.
  As a result, the drift of the charge distribution to the charge symmetric
 configuration was underestimated. This circumstance leaded to overestimation
 of the fusion factor $P_{\rm CN}$ in the former calculations of $\sigma_{\rm ER}$
 presented in Ref. \cite{NasirovPRC79}. We  discuss some details in Appendix.

\section{Conclusions}

In the framework of the combined DNS and advanced statistical
models, the ER excitation functions have been calculated for the
$^{48}$Ca+$^{249}$Bk reaction and the results are compared with
the experimental data given in  Ref. \cite{Ogan117}. The ER cross
section of the 4n-channel is  well described  while  the
3n-channel is described in  a satisfactory way,  in
both cases of the used  M\"oller and Nix
\cite{MolNix} and Muntian {\it et al.} \cite{Muntian03} mass
tables.

The capture, complete fusion and evaporation residue excitation
functions of the $^{50}$Ti+$^{252}$Cf and $^{54}$Cr+$^{248}$Cm
reactions, which could lead to  the synthesis of the
superheavy element $Z=120$,  have been calculated.
  The comparison of the results   show that at low  $E_{\rm \,c.m.}$  energies  the
 capture cross sections  of the $^{54}$Cr+$^{248}$Cm reaction
 are larger than  the  ones of the $^{50}$Ti+$^{249}$Cf reaction,
 while these cross sections become comparable at  higher energies corresponding
 to the 3n- and 4n-channel formations.
 The fusion cross section for the $^{50}$Ti+$^{249}$Cf reaction
 is   significantly larger than  that for the $^{54}$Cr+$^{248}$Cm reaction,
 though the former system has  a smaller number of
 neutrons than the latter one. The  stronger hindrance to complete fusion
 in  the case of the
 $^{54}$Cr+$^{248}$Cm reaction is connected with the  larger intrinsic fusion barrier
 $B^*_{\rm fus}$ and   smaller quasifission barrier $B_{\rm qf}$ than in the case of the
 $^{50}$Ti+$^{249}$Cf reaction.  In any case,  it appears in
the present study--when the M\"oller-Nix mass table is used--the maximum values of the excitation
function corresponding to the 3n-channel of the
evaporation residue formation for the $^{50}$Ti+$^{249}$Cf and
$^{54}$Cr+$^{248}$Cm reactions are  not higher than 0.1 and 0.07
pb, respectively,  while the maximum yield of residue for the
4n-channel (0.01 pb) for the reaction induced by $^{54}$Cr  is
higher than the one (0.004 pb) found for the reaction induced by
$^{50}$Ti.

\vspace{2cm}

\textbf{Acknowledgments}

A. K. Nasirov is grateful to the Istituto Nazionale  di Fisica Nucleare and
Department of Physics of the University of Messina for the support
received in the collaboration between the Dubna and Messina groups,
and he thanks the Russian Foundation for Basic Research for the partial
financial support in the  performance of this work.

\appendix*
\section*{Appendix}
The fusion factor $P_{\rm CN}(E,\ell)$ used in Eq. (\ref{Eq3})
shows the degree of hindrance
to complete fusion due to competition with quasifission.
The intense nucleon exchange between constituents of
DNS, which is formed at the capture of projectile by the target nucleus,
 can lead to formation of the compound nucleus or
 quasifission--DNS breaks down after intense mass transfer from the light
 constituent to the heavy one. For the heavy systems the hindrance
 to fusion increases and  $P_{\rm CN}(E,\ell)$ becomes very small
 in dependence on the mass asymmetry of the entrance channel.
\begin{figure}
\vspace*{0.0cm}
\begin{center}
\resizebox{0.80\textwidth}{!}{\includegraphics{{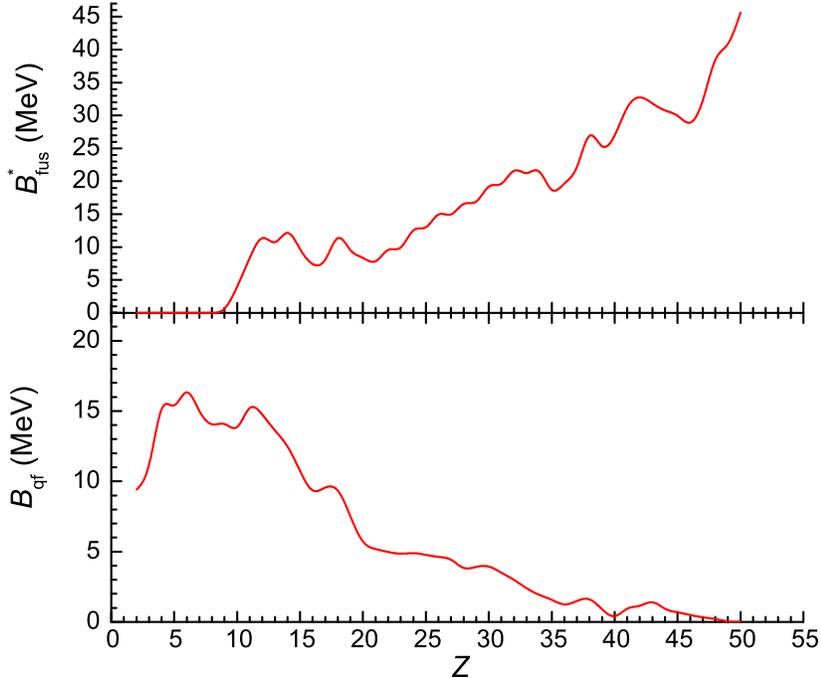}}}
\vspace*{-0.90 cm} \caption{\label{BfusBqf} (Color online)
The intrinsic fusion barrier (upper panel) and quasifission barrier (lower panel)
as functions of the charge asymmetry  for the $^{54}$Cr+$^{248}$Cm reaction.}
\end{center}
\end{figure}

The mass asymmetry degree of freedom may be fully or partially
equilibrated \cite{Bock}. Therefore, while DNS exists,
we have an ensemble \{$Z$\} of the DNS
configurations which contribute to the competition between
complete fusion and quasifission with probabilities \{$Y_Z$\}.

\begin{figure}
\vspace*{0.0cm}
\begin{center}
\resizebox{0.80\textwidth}{!}{\includegraphics{{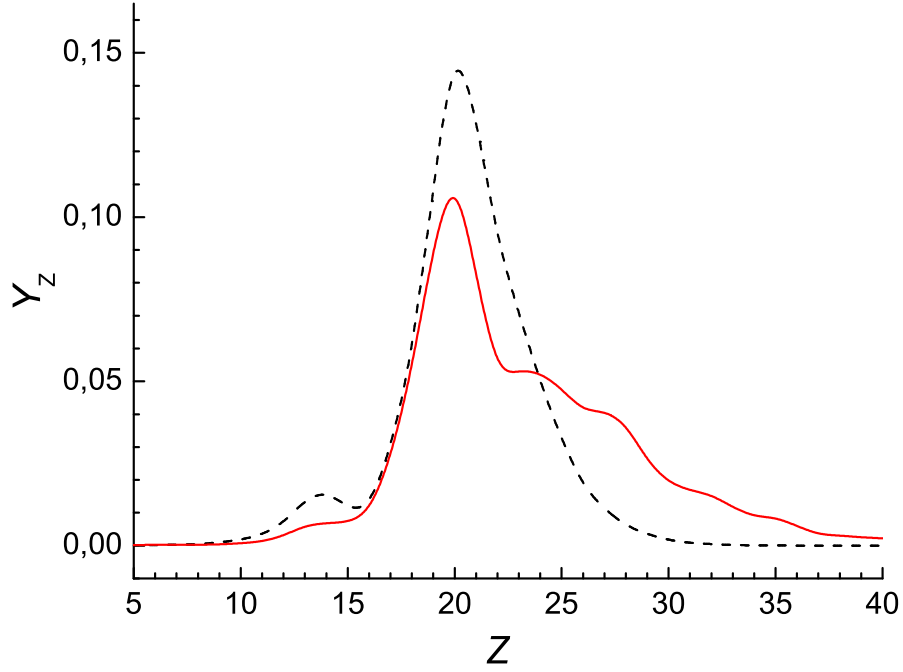}}}
\vspace*{-0.90 cm} \caption{\label{CompYz} (Color online)
Comparison  between the charge distributions in DNS
for the $^{54}$Cr+$^{248}$Cm reaction
 used in  Ref. \cite{NasirovPRC79} (dashed line) and in this work (solid line)  to
  calculate the complete fusion cross section.}
\end{center}
\end{figure}

The values of $B^*_{\rm fus}$ and $B_{\rm qf}$   are determined from the
landscape of the potential energy surface $U(A,Z;R,\ell)$.
 In Fig. \ref{BfusBqf} we present the result for the $^{54}$Cr+$^{248}$Cm reaction.

The $P_{\rm CN}$  factor depends on the charge
 distribution $Y_Z(E^*_{\rm DNS})$:
\begin{equation}
\label{pcne} P_{\rm CN}(E^*_{\rm DNS},\ell)=\sum_{Z_{sym}}^{Z_{max}}
Y_Z(E^*_{\rm DNS},\ell)P^{(Z)}_{\rm CN}(E^*_{\rm DNS},\ell),
\end{equation}
where $P^{(Z)}_{\rm CN}(E^*_{\rm DNS},\ell)$ is the fusion probability
for DNS having excitation energy $E^*_{\rm DNS}(Z)$ at charge asymmetry
$Z$. The method used to calculate $P^{(Z)}_{\rm CN}(E^*_{\rm DNS},\ell)$
is presented in Ref. \cite{anisEPJA34}.
The evolution of $Y_Z$  is calculated by solving the
transport master equation:
\begin{eqnarray}
\label{massdec}
\frac{\partial}{dt}Y_{Z}(E^*_Z,\ell,t)&=&\Delta^{(-)}_{Z+1}
Y_{Z+1}(E^*_Z,\ell,t)+
\Delta^{(+)}_{Z-1}  Y_{Z-1}(E^*_Z,\ell,t)\nonumber\\
&&-(\Delta^{(-)}_{Z}+\Delta^{(+)}_{Z}+\Lambda^{\rm qf}_{Z})
Y_{Z}(E^*_Z,\ell,t), \hspace*{0.1cm}\mbox{\rm for} \ Z=2,3,...,
Z_{tot}-2.
\end{eqnarray}
Here, the transition coefficients of multinucleon transfer  are
calculated as in Ref.~\cite{Jolos86}
\begin{eqnarray}
\label{delt} \Delta^{(\pm)}_{Z}=\frac{1}{\Delta t}
\sum\limits_{P,T}|g^{(Z)}_{PT}|^2 \ n^{(Z)}_{T,P}(t) \
(1-n^{(Z)}_{P,T}(t)) \
 \frac{\sin^2(\Delta t(\tilde\varepsilon_{P_Z}-
 \tilde\varepsilon_{T_Z})/2\hbar)}{(\tilde\varepsilon_{P_Z}-
 \tilde\varepsilon_{T_Z})^2/4},
\end{eqnarray}
where $\varepsilon_{i_Z}$ and  $n^{(Z)}_{i}(t)$ are the single-particle
energies and  occupation numbers of nucleons in the DNS fragments;
 the matrix elements $g_{PT}$  describe one-nucleon exchange
between the nuclei of DNS, and their values are calculated
microscopically using the expression obtained  in
Ref.~\cite{Adam92}. In the above-mentioned paper \cite{NasirovPRC79},
the diffusion of nucleons to the direction of the charge symmetric
 configuration of DNS was small due to the smallness of the $g_{PT}$ values
 which are determined by the meanfields of the interacting nuclei.
The radius coefficient $r^{\rm mfield}_0$ used in calculation
 of the nuclear meanfield was smaller in comparison with
 values of the radius coefficient $r^{\rm density}_0$ used in
 calculation of the nucleon density in nuclei. Therefore, when
 DNS is formed the distance between centers is determined
 by the minimum of the potential well of the nucleus-nucleus
 interaction  but at this distance the $g_{PT}$ values
 were small. This fact was not  adequately considered in our previous calculation
 presented in  the paper \cite{NasirovPRC79}.
 In Fig. \ref{CompYz}, we present the results of the charge distributions in DNS
 for the $^{54}$Cr+$^{248}$Cm reaction.

\end{document}